\documentclass[sigconf]{acmart} 

\usepackage{algorithm}
\usepackage[title]{appendix}
\usepackage{bm} 
\usepackage{color}
\usepackage{colortbl}
\usepackage{fancybox}
\usepackage{float}
\usepackage{lscape}
\usepackage{makecell}
\usepackage{multirow}
\usepackage{subcaption}
\usepackage{soul}
\usepackage{enumitem}

\usepackage{pifont}
\usepackage{listings}
\usepackage{xcolor}
\definecolor{darkgreen}{RGB}{0,120,0}

\newcommand{\system}{CalmReminder}

\usepackage{framed}
\usepackage{xcolor}

\definecolor{insightbg}{RGB}{250, 250, 250}
\definecolor{insightframe}{RGB}{150, 150, 150}
\definecolor{insightaccent}{RGB}{50, 140, 50}

\newsavebox{\insightboxcontent}
\newlength{\insightboxheight}

\newenvironment{insightbox}{%
    \begin{lrbox}{\insightboxcontent}%
    \begin{minipage}{0.87\linewidth}%
    \setlength{\parindent}{0pt}%
    \small
}{%
    \end{minipage}%
    \end{lrbox}%
    \setlength{\insightboxheight}{\dimexpr\ht\insightboxcontent+\dp\insightboxcontent+17pt\relax}%
    \par\vspace{6pt plus 2pt minus 2pt}\noindent
    \setlength{\fboxsep}{8pt}%
    \fcolorbox{insightframe}{insightbg}{%
        \kern-8.5pt%
        \kern12pt%
        \usebox{\insightboxcontent}%
        \kern6pt%
    }%
    \par\vspace{6pt plus 2pt minus 2pt}%
}
\newcommand{\tabref}[1]{Table~\ref{#1}}
\newcommand{\figref}[1]{Figure~\ref{#1}}
\newcommand{\secref}[1]{Section~\ref{#1}}

\newcolumntype{C}[1]{>{\centering\arraybackslash}m{#1}}

\newcommand{\etal}{\textit{et al.}}
\newcommand{\eg}{\textit{e.g.},~}
\newcommand{\ie}{\textit{i.e.},~}

\newcommand{\vs}{\textit{vs.}~}

\AtBeginDocument{%
  \providecommand\BibTeX{{%
    \normalfont B\kern-0.5em{\scshape i\kern-0.25em b}\kern-0.8em\TeX}}}

\setcopyright{acmlicensed}

\copyrightyear{2026}
\acmYear{2026}
\setcopyright{cc}
\setcctype{by}
\acmConference[CHI '26]{Proceedings of the 2026 CHI Conference on Human Factors in Computing Systems}{April 13--17, 2026}{Barcelona, Spain}
\acmBooktitle{Proceedings of the 2026 CHI Conference on Human Factors in Computing Systems (CHI '26), April 13--17, 2026, Barcelona, Spain}
\acmPrice{}
\acmDOI{10.1145/3772318.3791871}
\acmISBN{979-8-4007-2278-3/2026/04}

\begin{document}

\title[]{\system: A Design Probe for Parental Engagement with Children with Hyperactivity, Augmented by Real-Time Motion Sensing with a Watch}

\author{Riku Arakawa}
\authornote{Both authors contributed equally to this research.}
\orcid{0000-0001-7868-4754}
\affiliation{%
  \institution{Human-Computer Interaction Institute\\Carnegie Mellon University}
  \city{Pittsburgh}
  \state{Pennsylvania}
  \country{USA}
}
\email{rarakawa@cs.cmu.edu}

\author{Shreya Bali}
\authornotemark[1]
\orcid{0000-0002-6005-409X}
\affiliation{%
  \institution{Human-Computer Interaction Institute\\Carnegie Mellon University}
  \city{Pittsburgh}
  \state{Pennsylvania}
  \country{USA}
}
\email{sbali@andrew.cmu.edu}

\author{Anupama Sitaraman}
\orcid{0009-0009-1790-5846}
\affiliation{%
  \institution{Carnegie Mellon University}
  \city{Pittsburgh}
  \state{Pennsylvania}
  \country{USA}
}
\email{asitaram@andrew.cmu.edu}

\author{Woosuk Seo}
\orcid{0000-0002-1316-4341}
\affiliation{%
  \institution{Emergency Medicine\\Yale University}
  \city{New Haven}
  \state{Connecticut}
  \country{USA}
}
\email{woosuk.seo@yale.edu}

\author{Sam Shaaban}
\orcid{0000-0001-8273-1517}
\affiliation{%
  \institution{NuRelm}
  \city{Pittsburgh}
  \state{Pennsylvania}
  \country{USA}
}
\email{sam@nurelm.com}

\author{Oliver Lindhiem}
\orcid{0000-0002-1012-9102}
\affiliation{%
  \institution{Department Psychiatry\\University of Pittsburgh School of Medicine}
  \city{Pittsburgh}
  \state{Pennsylvania}
  \country{USA}
}
\email{lindhiemoj@upmc.edu}

\author{Traci M. Kennedy}
\orcid{0000-0001-9865-3533}
\affiliation{%
  \institution{Department Psychiatry\\University of Pittsburgh School of Medicine}
  \city{Pittsburgh}
  \state{Pennsylvania}
  \country{USA}
}
\email{kennedytm2@upmc.edu}

\author{Mayank Goel}
\orcid{0000-0003-1237-7545}
\affiliation{%
  \institution{School of Computer Science\\Carnegie Mellon University}
  \city{Pittsburgh}
  \state{Pennsylvania}
  \country{USA}
}
\email{mayankgoel@cmu.edu}

\renewcommand{\shortauthors}{Arakawa and Bali et al.}


\begin{abstract} 
Families raising children with ADHD often experience heightened stress and reactive parenting.
While digital interventions promise personalization, many remain one-size-fits-all and fail to reflect parents' lived practices.
We present \system{}, a watch-based system that detects children's calm moments and delivers just-in-time prompts to parents. 
Through a four-week deployment with 16 families (twelve completed) of children with ADHD, we compared notification strategies ranging from hourly to random to only when the child was inferred to be calm.
Our sensing-based notifications were frequently perceived as arriving during calm moments.
More importantly, parents adopted the system in diverse ways: using notifications for praise, mindfulness, activity planning, or conversation. 
These findings show that parents are not passive recipients but active designers, reshaping interventions to fit their parenting styles.
We contribute a calm detection pipeline, empirical insights into families' flexible appropriation of notifications, and design implications for intervention systems that foster agency.
\end{abstract}



\maketitle

\section{Introduction}
\label{sec:intro}

\begin{figure*}[t]
    \centering
    \includegraphics[width=0.95\linewidth]{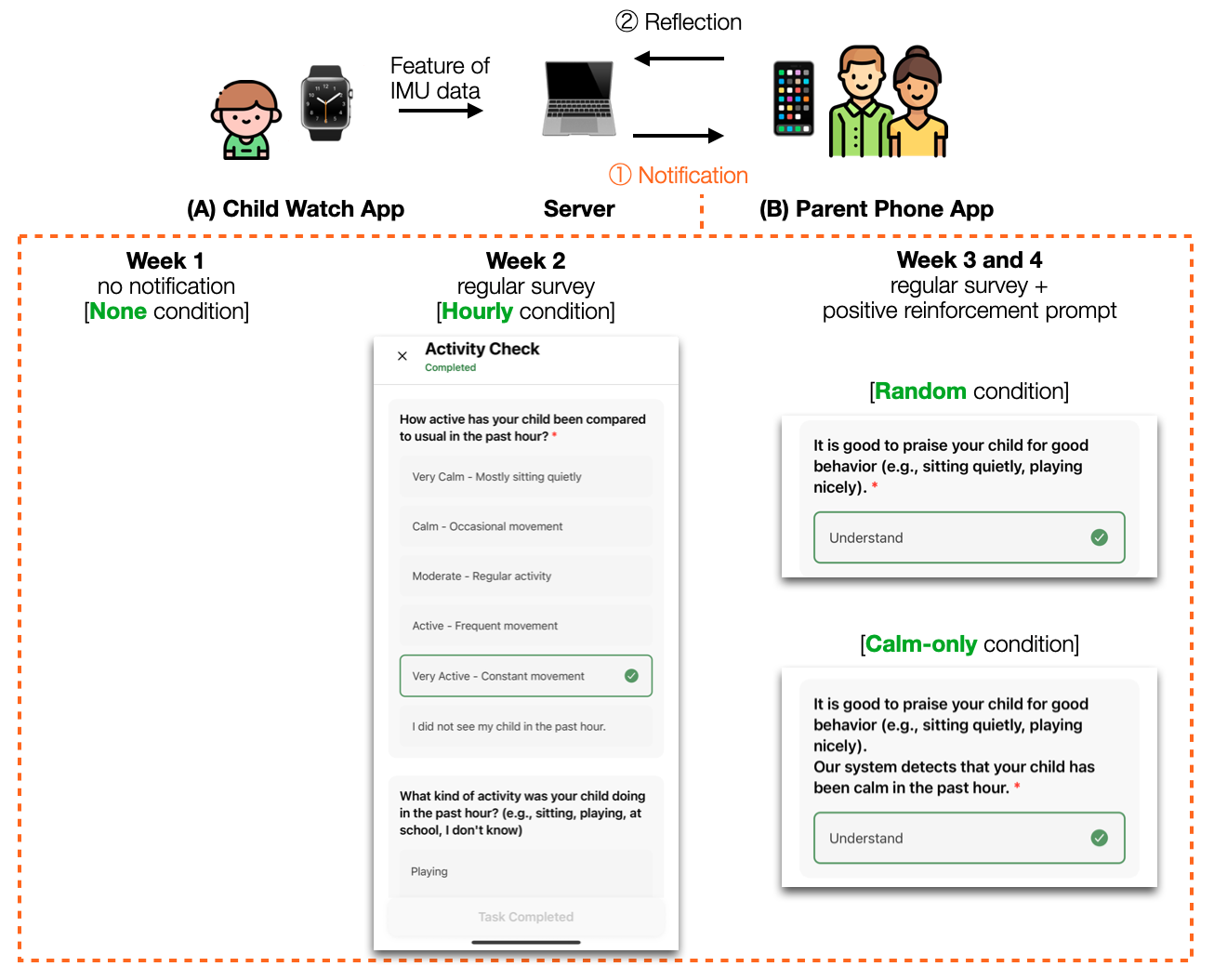}
    \caption{Overview of the developed \system{} system. A child with hyperactivity wears a smartwatch with our custom app (A), which continuously sends IMU data to our server. The server periodically sends notifications to the phone app (B) on the parent's phone \ding{192}, where they respond to the delivered reflection prompts \ding{193}. We conducted a four-week study to explore different notification strategies: none, hourly, random, and calm-only.}
    \Description{High-level system diagram describing the \system{} system. In the upper part of the diagram, 3 entities are shown: There is a child with a smartwatch and a caption reading "(A) Child Watch App", there is an image of parents with a phone and a caption reading "(B) Parent Phone App, and there is an image of a computer with the caption "Server".
    There are labeled arrows connecting these three entities. There is an arrow from the child to the server labeled "IMU Data". There is an arrow from the server to the parents labeled "Notification". There is an arrow from the parents to the server reading "Reflection". In the bottom part of the figure, the "Notification" is described further. Three notification conditions are described, with example screenshots of what is shown in the parent phone app. From left to right, there are labels for the notification for "week 1", "week 2", and "week 3 and 4". For "week 1", there is no notification and the condition is "None". For "week 2", there is a regular survey and the condition is "Hourly". The screenshot of the survey is shown. For "week 3 and 4", there is a regular survey and a positive reinforcement prompt. There is also a "Random" condition and a "Calm-only" condition. Screenshots of the positive reinforcement prompt are shown for both the "Random" and "Calm-only" conditions.}
    \label{fig:system-overview}
\end{figure*}

\begin{figure*}[t]
    \centering
    \includegraphics[width=0.7\linewidth]{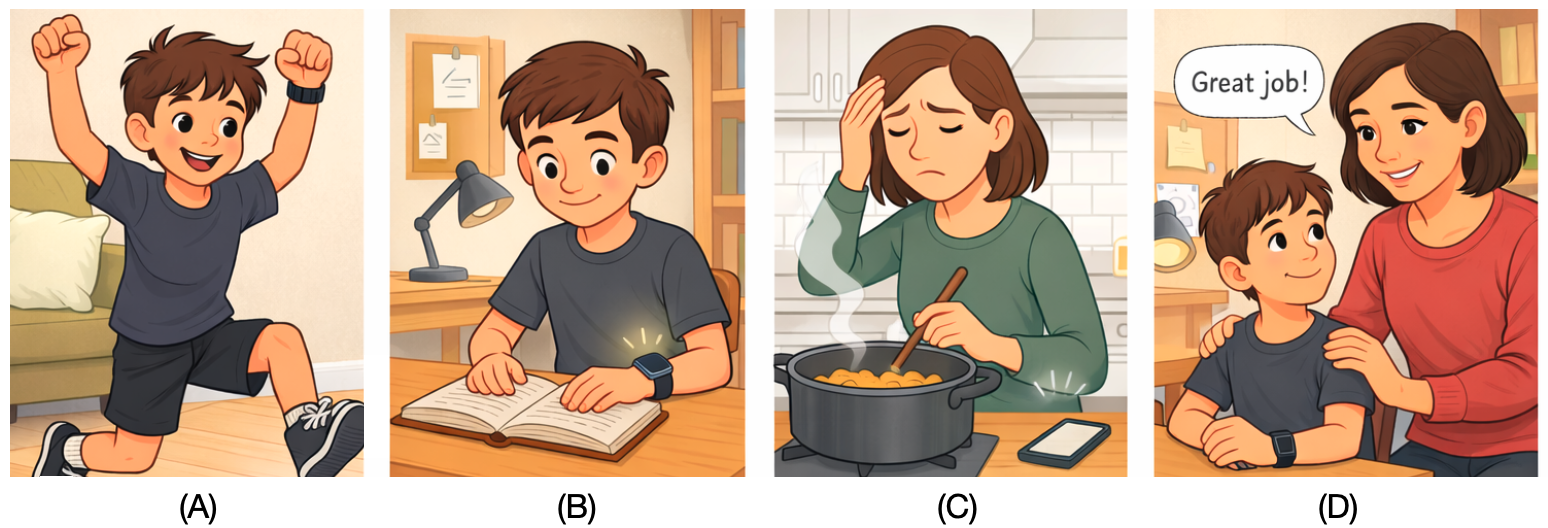}
    \caption{Example scenario with \system{} as observed in our four-week in-home study. A child with hyperactivity wears a smartwatch (A) that detects calm moments through motion sensing (B). When the parent, occupied with daily tasks, receives a notification, they are reminded to offer positive reinforcement (C). This reinforcement strengthens the parent-child relationship (D). Our study further revealed diverse ways families appropriated \system{}'s features based on their parental values and practices. Each picture was generated by ChatGPT-5.
    }
    \Description{Four sepia-toned illustrations showing an example scenario with CalmReminder.
    (A) A young boy with hyperactivity is shown leaping with excitement, wearing a smartwatch.
    (B) The same boy sits calmly at a desk, reading a book while the smartwatch glows to indicate detection of calm behavior.
    (C) His mother is in the kitchen, busy cooking at the stove, while her phone on the counter glows with a notification.
    (D) The mother praises the boy, placing a supportive hand on his shoulder as he smiles, reinforcing their bond.}
    \label{fig:envisioned-scenario}
\end{figure*}

Attention-Deficit/Hyperactivity Disorder (ADHD) is a neurodevelopmental disorder characterized by heightened levels of inattentiveness, impulsivity, and hyperactivity, with symptoms often emerging as early as four years of age.
Globally, ADHD affects approximately 5\% of children and adolescents~\cite{song2021prevalence,edition2013diagnostic,polanczyk2007worldwide}.
Among school-aged children diagnosed with ADHD, roughly 55\% present with prominent hyperactive symptoms~\cite{willcutt2012prevalence}.
Families raising children with hyperactivity face challenges in everyday life as children's hyperactive episodes often disrupt daily routines, elevate parental stress, and strain family relationships~\cite{Whalen2011, Johnston2001}.
Conversely, parental stress and negative interactions can exacerbate the child's behavioral difficulties, creating feedback loops that are difficult to break~\cite{Whalen2009, GERDES_HOZA_PELHAM_2003}.
These dynamics highlight the need for strategies that support parents in maintaining constructive interactions at home and in recognizing opportunities for positive engagement~\cite{Li2018ASmartphone}.

HCI research has investigated how digital technologies can scaffold behavior management and family well-being in ADHD contexts.
Existing systems range from self-regulation tools that support children directly (\eg focus aids, emotion tracking)~\cite{DBLP:conf/assets/Dibia16,DBLP:journals/tochi/AnkrahCSTBSLH23}, to co-regulation approaches that engage parents and caregivers through collaborative data sharing and daily routine support~\cite{Paidipati2020,DBLP:conf/chi/SonneMMOG16,DBLP:conf/chi/SilvaCMBBBESLH23}.
These results and insights across decades of social learning and parent training research suggest that timely parental engagement—particularly moments of positive reinforcement—is crucial for reducing reactive parenting patterns~\cite{Leijten2019,skinner2019behavior,barkley2013defiant,kazdin2021kazdin}.
Operant conditioning and evidence-based parent training programs emphasize that reinforcing desired behaviors (such as calm, compliant behavior) and strategically withdrawing attention from undesired behaviors is more effective than punishing problematic behavior~\cite{Eyberg1988Parent-Child,KAZDIN1997Parent,patterson1982coercive,webster2018incredible}.
Yet, a well-documented challenge is that calm, desired behaviors are harder for parents to notice in the moment than disruptive behaviors that naturally capture attention~\cite{patterson1982coercive}.
As a result, parents often overlook everyday moments of desirable behavior, even though attending to and reinforcing such moments plays a central role in strengthening positive parent-child interactions and promoting behavioral change.

From these insights, we argue that technologies for ADHD and family support should help parents recognize and reflect on children's calm moments so they can reinforce them rather than monitoring or flagging problematic behavior.
However, prior systems have not examined how technology might identify low-hyperactivity periods in real time and translate them into actionable prompts for parents.
In particular, it remains unclear how such prompts might shape caregivers' in-the-moment perceptions of their child's behavior and how these perceptions evolve into daily reflection.

In this paper, we present \textit{\system{}} as a design probe to examine how wearable sensing and notification design can shape parental engagement for parents raising children with ADHD (\figref{fig:system-overview}).
By configuring different notification strategies, we investigated how real-time sensing could be translated into moments of parent-child interaction.
The system continuously collected motion data (IMU) from a child's smartwatch and surfaced periods of low hyperactivity as lightweight notifications to parents' smartphones.
These timely prompts were designed to help parents notice and reflect on their child's behavior in vivo, and to support them in positively reinforcing calm, desired behavior that might otherwise go unnoticed.
Through this probe, we surface broader questions about how real-time sensing and lightweight prompts can contribute to the relational dynamics of family life, rather than simply delivering behavioral updates.

We deployed \system{} in a four-week in-home study with 16 (twelve completed) families of children diagnosed with ADHD. 
Through this deployment, families experienced four intraday notification conditions (each for a week):
\begin{itemize}
    \item \textbf{None}: No notifications. Only passive data collection.
    \item \textbf{Hourly}: Regular notifications set to once per hour. This condition also provided data for personalization of our machine-learning model.
    \item \textbf{Random}: Five notifications in a day at random times.
    \item \textbf{Calm-only}: Up to five notifications when the sensing model on the watch inferred the child to be calm.
\end{itemize}
Intraday notifications were delivered at most once per hour. 
In addition, the random and calm-only conditions included reminders prompting parents to positively reinforce their child's good behavior, as shown in~\figref{fig:system-overview}.

We examined both the accuracy of our sensing model in detecting calm moments and how different notification strategies shaped parental awareness and reflection. 
The calm-moment detection algorithm delivered notifications at times that parents themselves perceived as calm; among the notifications to which parents responded, 78.0\% were reported as actually calm.

Rather than prescribing a single ``correct'' way to use \system{}, we intentionally left feature use open-ended as a design probe.
This revealed a clear pattern during deployment: parents did not passively adopt or reject the system as designed, but actively functioned as active designers of their own interventions. They selectively attended to different features, repurposed notifications to fit diverse goals (mindfulness, conversation starters, activity scheduling), and integrated sensing capabilities into their existing parenting practices. This flexible appropriation challenges the dominant paradigm of designing singular interventions to be evaluated for ``effectiveness,'' instead suggesting that configurable features enable families to construct personalized strategies suited to their unique contexts.
Building on these insights, we outline design implications for Just-in-Time Adaptive Interventions (JITAIs)~\cite{NahumShani2017JITAIs} grounded in behavior-based prompting in family settings, and discuss how such interventions can shape families' dyadic reflection and reinforcement practices through the lens of Dyadic Illness Management~\cite{Lyons2018Theory}.
Instead of one-size-fits-all interventions, our goal is to position technology as a configurable resource that families can integrate into their own styles and relational routines (\figref{fig:envisioned-scenario}).

Our work contributes to the following:
\begin{enumerate}
    \item \system{}, a real-time system that leverages wearable sensing to detect children's calm moments and deliver just-in-time prompts to parents for reflection, with code available at \url{https://github.com/cmusmashlab/calmreminder}. 
    \item Results from a four-week in-home deployment with the completed 12 families of children with hyperactivity, comparing different notification strategies, which confirmed the accurate calm-moment detection of our system and parents' various ways to adopt and adapt \system{}'s features.
    \item Design implications for flexible and personalized interventions that support parental reflection and allow families to configure timing and engagement styles to their own management strategies.
\end{enumerate}

\section{Related Work}
\label{sec:rw}

Our work builds on and bridges three strands of prior research: (1) family systems perspectives and parent-based behavioral interventions for pediatric ADHD, (2) digital technologies that support self-regulation and co-regulation in families, and (3) behavioral sensing for hyperactivity.

\subsection{Family Systems and Parent-Based Behavioral Interventions for ADHD}
\label{sec:rw-family}

Prior work on interventions for children with ADHD highlight the central role of the family environment and the parenting practices that shape children's behavioral trajectories.
We situate our work within these perspectives, drawing on Family Systems Theory (FST) and well-established parent-based behavioral interventions.
FST posits that family members function as an interconnected system where changes in one member's behavior ripple throughout the entire unit~\cite{Minuchin2009}.
It has been used to better understand family interactions and to inform the design of technology, such as the integration of family-centered robots into homes~\cite{cagiltay2023family}.
In families of children with ADHD, these dynamics manifest as \textit{bidirectional influences}~\cite{Johnston2001}: a child's hyperactive episodes can elevate parental stress~\cite{Whalen2011}, while parental stress can exacerbate the child's regulatory difficulties~\cite{Whalen2009, Bilgi2016}.
Crucially, these cycles can be either detrimental or beneficial.

Research demonstrates that parent-based behavioral interventions are highly effective in improving behavioral outcomes in children with ADHD, and that having established such interventions can increase a parent's sense of competence and lower their stress~\cite{DEKKERS2022478}.
In this regard, Leijten~\etal~\cite{Leijten2019} suggested that timely \textit{positive and active} reinforcement of calm, compliant behavior not only reduces disruptive child behavior but also enhances parental mood and confidence.
However, the special conditions of parenting a child with ADHD often lead parents to develop more \textit{negative and reactive} parenting patterns~\cite{Johnston2001, GERDES_HOZA_PELHAM_2003}.
This negativity bias can cause parents to systematically overlook opportunities for positive engagement, particularly during periods when their child exhibits calm or appropriate behavior, precisely the moments when positive reinforcement would be effective and relationship-building~\cite{Lench2011Exasperating}.
Further, while research has established the critical importance of positive engagement, no studies have empirically examined or compared different technology-mediated strategies for encouraging parental positive reinforcement with children who have hyperactivity.
Our work addresses this gap by operationalizing these principles through real-time, sensing-driven prompts that highlight calm moments in everyday life and by examining how such prompts fit within families' existing interaction patterns.

\subsection{Digital Technologies for ADHD: Self-Regulation and Co-Regulation}

ADHD is a chronic lifelong condition that requires active self-management and ongoing support from healthcare providers, families, and caregivers.
Traditional psychosocial treatments for ADHD face persistent challenges in ensuring that therapeutic gains generalize beyond structured clinical settings into the complexity of everyday life~\cite{Abikoff2009ADHD,Mikami2017Social}. 
This difficulty highlights the importance of interventions that can provide timely, context-sensitive support in situ.
A growing body of HCI research has, therefore, explored how digital technologies can support individuals with ADHD. 

\textbf{Self-Regulation approaches:} One prominent area focuses on self-regulation tools that help patients manage ADHD symptoms independently.
For example, Dibia~\etal~\cite{DBLP:conf/assets/Dibia16} designed a smartwatch application ``FOQUS''  to help users extend focus time and tools to reduce anxiety and stress.
Building on this individual-focused approach, Ankrah~\etal~\cite{DBLP:journals/tochi/AnkrahCSTBSLH23} investigated how children with ADHD understand and interpret their own smartwatch-derived activity data, revealing important insights about children's capacity for self-reflection and data literacy in ADHD self-management. 

\textbf{Co-regulation approaches:} For children with ADHD, however, research has long identified that family and caregivers play a crucial role in helping manage the condition~\cite{Paidipati2020, Sin2022}. 
Consequently, a second line of research emphasizes co-regulation - collaborative approaches that leverage family dynamics to support the child's symptom management. 
Cibrian~\etal~\cite{DBLP:conf/chi/CibrianLTGSH20} created design workshops to help understand the balance between co-regulation and self-regulation and identify opportunities for children with ADHD.
This family-centered perspective has generated diverse technological interventions spanning multiple domains: prior work explored emotional development tools that help children with ADHD articulate emotions and engage in structured reflection~\cite{DBLP:conf/acmidc/StefanidiWWSRN24, DBLP:journals/pacmhci/StefanidiWCWNMRN25}, while collaborative data sharing systems enable families to jointly interpret activity data for learning and skill development~\cite{DBLP:conf/chi/SilvaCBBMMVSLHE24}.
Similarly, routine-focused interventions target specific behavioral challenges, ranging from bedtime management systems~\cite{DBLP:conf/chi/SonneMMOG16, DBLP:conf/acmidc/SonneMMOG16} to gamified daily activity planning tools that use point-based incentive structures to foster family collaboration~\cite{DBLP:conf/chi/SilvaCMBBBESLH23}. 

In this line of research, Pina~\etal~\cite{DBLP:conf/ph/PinaRRJHC14} demonstrated that mobile technology can deliver parenting strategy reminders by detecting parental stress, showing that timely interventions help parents implement behavioral strategies more effectively at critical moments.
Their findings revealed that parents often struggle to recall and apply these strategies precisely when stress is highest, underscoring the importance of not only \textit{what} support is provided but also \textit{when} it is delivered.
Building on this insight, our work explores a novel approach to intervention timing that leverages ubiquitous sensing to account for children's behavioral states, thereby creating opportunities for positive reflection and reinforcement (as discussed in~\secref{sec:rw-family}) and ultimately fostering more constructive parent-child interactions.

\subsection{Behavioral Sensing for Hyperactivity}

Researchers have explored automating the detection of hyperactivity in children~\cite{DBLP:journals/cbm/LohPBPMA22,Wang2022AReview} using diverse modalities such as MRI, EEG, and motion sensors. 
Among these, motion-sensor–based approaches are most prevalent, since hyperactivity often manifests as frequent and excessive body movement~\cite{Lin2020Quantitative,Earnest2020Actigraph,Lindhiem2022Objective,DBLP:journals/imwut/ArakawaAMTSLG23}. 
For example, Lin~\etal~\cite{Lin2020Quantitative} tracked children's arm movements during seated classroom activities using smartwatches, reporting significant differences in gyroscope and accelerometer zero-crossing rates between children with and without hyperactivity. 
Arakawa~\etal~\cite{DBLP:journals/imwut/ArakawaAMTSLG23} correlated smartwatch-derived activity levels in daily life with hyperactivity risk, suggesting that IMU sensors can provide a reliable indicator of hyperactivity. 
Their study also showed that incorporating parent-reported behavioral input, provided through end-of-day surveys, helps contextualize sensor data to improve the model estimation.

These sensing-based approaches show promise, but they are not yet well integrated into parent-child interactions.
First, many rely on diagnostic labels (\eg hyperactive \vs non-hyperactive) as training targets for machine-learning models, without incorporating parents' perceptions of their child's activity in daily life.
Second, as Arakawa~\etal's study~\cite{DBLP:journals/imwut/ArakawaAMTSLG23} suggests, while personalization based on parental input is valuable, collecting this input through end-of-day surveys often adds extra burden.
We reimagine this process through real-time notifications asking parents about their child's activity level while simultaneously offering opportunities for reflection.
In doing so, the sensing model and notification timing will be personalized to each parent's perceptions, offering insights into their children's behavior.

\section{System}
\label{sec:system}

We developed a custom probe for this study, which we call \system{}.
The overview is presented in~\figref{fig:system-overview}.
The system consists of three core components: a smartwatch app worn by the child, a mobile app used by parents to receive notifications and respond to surveys, and a web-based server dashboard used by experimenters to manage the study.

\subsection{Design Considerations}
\label{sec:system-design}

In designing our system, we aimed to sense children's behavior with minimal device installation and to share insights with parents in real-time, while also fostering reflection that could strengthen parent-child relationships over time.
We selected the smartwatch as the sensing platform because of its ubiquity and relatively low risk of privacy invasion.
Prior research has also highlighted the device's acceptability for children~\cite{Thompson2025Exploring,Cibrian2021Parents}.

Importantly, our design avoids positioning the watch as a surveillance tool directed at the child~\cite{DBLP:conf/acmidc/KumarOLBCCV23}.
To achieve this, we restricted the visibility of sensor data: parents never see raw motion values.
Instead, the sensed information is abstracted into notification timing and carefully crafted message content. 
This ensures that the system supports awareness and reflection without exposing sensitive data or encouraging parental over-monitoring.

A key intervention strategy we introduced is the delivery of notifications specifically during calm moments detected through motion sensing.
Below, we provide further rationale based on well-established psychological principles of behavioral learning and the resulting evidence-based therapeutic models.

\subsubsection{Why Calm Moments?}
\label{sec:system-design-why}

As discussed in Section~\ref{sec:rw-family}, family systems perspectives and parent training research highlight the central role of positive reinforcement and its timing in parent-based behavioral intervention for children with ADHD.
Here, the theoretical rationale for \system{} to alert caregivers to moments of calm (rather than hyperactivity) lies in social learning principles.
Most notably, operant conditioning~\cite{skinner2019behavior}, one of the most well-established evidence-based laws of human behavior, holds that behaviors that are rewarded become reinforced, whereas those that are not rewarded eventually wane.
These dual principles have led to several parent training programs that have demonstrated decades of evidence for their efficacy in reducing children's problematic behavior, including oppositional behavior as well as hyperactivity~\cite{barkley2013defiant,Eyberg1988Parent-Child,KAZDIN1997Parent,patterson1982coercive,webster2018incredible}.
Crucially, two key ingredients across these efficacious parenting techniques include: (1) positively reinforcing desired behavior, such as moments of calm and compliance, through parenting behaviors that are rewarding to children, such as verbal praise, positive attention, and sometimes tangible rewards or privileges, and (2) intentionally ignoring undesired behavior, such as moments of hyperactivity and noncompliance, which removes the parental attention (even if negative) to these behaviors that typically reinforce them~\cite{dishion1992experimental,Patterson1994AFunctional,KAZDIN1997Parent}.
Indeed, a vast literature documents the greater effectiveness of reinforcing desired behaviors relative to punishing undesired behaviors~\cite{KAZDIN1997Parent,kazdin2021kazdin}.
Further, ample evidence shows it is much harder for parents to notice calm, compliant, desired behavior than it is to notice hyperactive, noncompliant, undesired behavior that more naturally attracts attention~\cite{patterson1982coercive}.
For these reasons, \system{} was designed to provide a real-time nudge in response to children's moments of low hyperactivity to help parents reflect and positively reinforce this behavior rather than having it go unnoticed and unreinforced.

\subsection{Components}

\paragraph{Watch App}

Our custom app runs on an Apple Watch (Series 7 or later) and continuously collects IMU sensor data throughout the day (or as long as the battery lasts) (\figref{fig:system-overview}A). 
It computes a motion energy value in real time, defined as the root mean square of the acceleration signal over a 5-minute window.
When an internet connection is available, this value is uploaded to the server continuously.
We tested and confirmed that the app could operate for approximately 8–10 hours on a single charge. To minimize accidental interactions, a four-digit PIN is required to stop recording, ensuring intentional control by parents.
At the end of each day, parents are instructed to upload data via a one-tap button.
Uploads are resumable and designed to tolerate intermittent connectivity.

\paragraph{Phone App}

Another custom app is installed on the parent's personal smartphone and serves as the primary interface for notification and survey interactions (\figref{fig:system-overview}B).
Throughout the day, the app delivers intraday notifications with brief prompts that ask parents to reflect on their child's recent behavior.
In addition, there are end-of-day and end-of-week surveys. 
Details of these surveys are provided later in~\secref{sec:study-design}. 
The app visually distinguishes pending, completed, and expired tasks, and all surveys are designed to be lightweight (typically under one minute).
Notifications are time-sensitive to promote in-the-moment reflection, with an expiration time of 30 minutes for the intraday, 12 hours for the end-of-day, and 2 days for the end-of-week surveys.

\paragraph{Experimenter Dashboard}

To manage participants and notification policies, we developed a web-based experimenter dashboard.
The dashboard allows experimenters to register participants, monitor watch data uploads, track survey completion, and switch between notification strategies.
It also provides real-time data on device status (\eg battery level, connectivity), as well as logging of incoming data streams from both the watch and the phone app.

\subsection{Model to Detect Calm Moments and Notification Strategy}
\label{sec:system-algorithm}

\system{} has a feature to send an intraday notification when the child is detected to be calm.
We used IMU data from the smartwatch to estimate the level of activeness as it might be perceived by the parent.
Specifically, we used the energy of the acceleration as a feature, \ie the squared magnitude of the accelerometer signal integrated over time and averaged across the three axes.
Prior work~\cite{DBLP:journals/imwut/ArakawaAMTSLG23} indicated that this energy value correlates with the child's activity level in general.
For each participant, we trained a simple linear regression model mapping IMU-derived energy values to perceived parental activity levels, using scikit-learn's default implementation~\cite{DBLP:journals/corr/abs-1201-0490}. 
Here, the input corresponds to the average energy value during the one-hour period prior to when the parent completed the questionnaire.

Training data for the model was collected in the early days of the study through intraday surveys (\ie the hourly condition in Week 2, as we describe in the next section).
During this time, parents rated their child's activity level on a five-point Likert scale (1 = Very Calm, 5 = Very Active).
To avoid overloading parents, we enforced a minimum one-hour interval between notifications. The server then applied the trained model to estimate parental perception and determined whether to send a notification to the parent's phone. 
In one condition in the user study, \system{} triggered a notification when the model's output is less than 3, prompting the parents to reflect on their child's behavior during calm moments.

Our approach uses only IMU data as input, in contrast to studies like LemurDx~\cite{DBLP:journals/imwut/ArakawaAMTSLG23,Lindhiem2022Objective} which emphasize contextualization to interpret children's activity (\eg distinguishing play from dysregulation).
We hypothesize that calm moments are semantically more stable and that low IMU activity usually provides sufficient indication of calm behavior.
Unlike active moments, which require extensive context labeling to disambiguate, calm moments can be identified without fine-grained contextual information.

This lightweight design works well within smartwatch battery constraints and can run directly on-device. Unlike computationally heavy models suggested in prior work, our approach avoids the need to continuously stream IMU data to remote servers, which would require stable Internet connectivity that could not be guaranteed in our in-the-wild deployment.
To balance these trade-offs, \system{} computed lightweight energy features from IMU data locally on the smartwatch and uploaded them to the server every five minutes.
Importantly, our results show that this energy-based method effectively modeled parents' perceptions in the wild.

\section{User Study}
\label{sec:study}

We conducted a four-week within-subjects field study employing four intraday notification timing strategies: none, hourly, random, and calm-only notifications.
The study aimed to assess the feasibility and user experience of using child motion data to promote parental reflection and engagement.

\subsection{Research Questions}
\label{sec:study-rqs}

Through this user study, we wished to explore the following research questions: 
\begin{description}
    \item[RQ1:] Can real-time motion sensing reliably reflect parents' perceptions of when their child is calm?
    \item[RQ2:]  How do families engage with and make sense of reflective prompts embedded in everyday routines?
    \item[RQ3:] How do different prompting strategies influence parents' experience with our system?
\end{description}

\subsection{Participants} \label{sec:study-participants}

\begin{figure*}[t]
    \centering
    \includegraphics[width=0.85\linewidth]{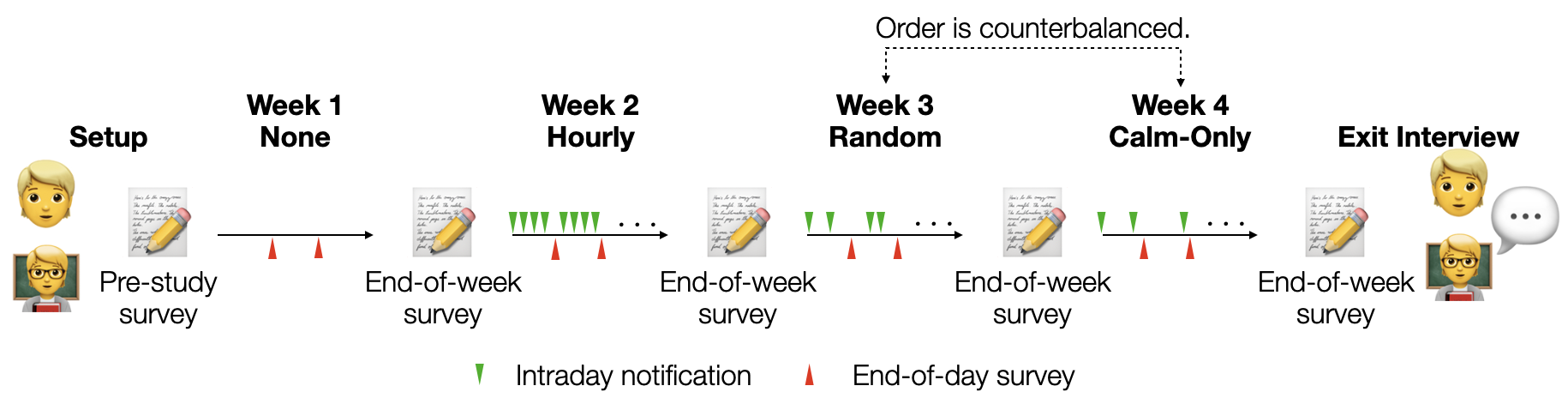}
    \caption{Overview of our four-week study involving different intraday notification strategies: none, hourly, random, and calm-only.}
    \Description{The overview of the 4-week study timeline, beginning with "Setup" and ending with "Exit Interview". From left to right, there are 6 captions: "Setup", "Week 1 None", "Week 2 Hourly", "Week 3 Random" "Week 4 Calm-Only" and "Exit Interview". Under "Setup", there is an image of a person and another image of a researcher, as well as an image of a paper and pencil labeled "Pre-study survey". "Week 1 None", "Week 2 Hourly", "Week 3 Random" and "Week 4 Calm-Only" have arrows underneath them displaying the notifications that the participant will experience during the week. At the end of each arrow, there is an image of a pencil and paper with a label "End-of-Week Survey". 
    Some of the arrows contain symbols to indicate the type of notification that occurs during the specific condition. There are green symbols indicating intraday notifications, and red symbols indicating end-of-day notifications. Under "Week 2 Hourly" the arrow has many green symbols, and sparse red symbols. These symbols also appear on the "Week 3 Random" and "Week 4 Calm-Only" arrows, though there are fewer green symbols. Finally, there is a double-headed arrow above "Week 3 Random" and "Week 4 Calm-Only" that indicates that Week 3 and Week 4 conditions are counterbalanced. }
    \label{fig:study-overview}
\end{figure*}

We recruited 16 parent-child pairs, requiring that children have a formal ADHD diagnosis and be aged 6–12 years.
The children had a mean age of 8.5 years, 12 of them were male. All parents were mothers.
Participants received \$75~USD compensation upon study completion.
The study was approved by the institute IRB [Hidden for Anonymity].

We used a pre-study survey to gauge the level of hyperactivity for each child. Specifically, we used five questions from the NICHQ-Vanderbilt-Assessment-Scale (VADRS)~\cite{VADRS} questionnaire. These questions were presented as a Likert scale where parents were asked to rate their child's behavior on a scale of 0 (never) to 3 (very often). Specifically, we asked the following questions: (1) Fidgets with hands or feet or squirms in seat, (2) Leaves seat when remaining seated is expected, (3) Runs about or climbs excessively in situations when remaining seated is expected, (4) Has difficulty playing or engaging in leisure/play activities quietly, and (5) Is ``on the go'' or often acts as if ``driven by a motor''.
Using these questions, we calculate a subsection of the total VADRS score.
We find that the majority of the children in the study (12/16) scored more than 10 points of the maximum possible point value of 15, indicating that the children in our study are highly hyperactive.

In addition, parents employed a variety of pre-existing management strategies prior to the study. We grouped participants who completed the study into four distinct categories to contextualize the intervention's adoption. These categories reflect the parents' primary focus and the mechanism of their established routine. We lack this information for three out of the four dropout participants, as they refused an interview.
The single dropped-out participant who agreed to an interview had no set strategy.

\begin{itemize}
    \item \textbf{Parent-Child Relationship-focused strategies} (P1, P3, P4, P5, P9, P10): Strategies focusing on structured communication with the child and improving the parent-child relationship, informed by expert advice (\eg parent-child interaction therapy).
    \item \textbf{Antecedent-focused strategies} (P2, P6, P8, P12): Prioritizing physical exertion and activities to regulate physiological energy levels
    \item \textbf{Consequence-based / contingency management strategies} (P7): Established token economies or privilege systems (\eg gaming/bikes) contingent on 'good behavior' and taking away those privileges for 'bad behavior'
    \item \textbf{No set strategy} (P11): One parent reported that they had no established strategy to manage their child's ADHD.
\end{itemize}

\subsubsection{Limitation}
We note that our participant pool is biased towards male children.
This is likely because male children are almost twice as likely to be diagnosed with ADHD~\cite{cdc_adhd_stats}.
The fact that the lower likelihood of female children having ADHD diagnoses made fewer female children eligible for our study.

\subsection{Design}
\label{sec:study-design}

We employed a within-subjects repeated measures design over four weeks to explore different notification strategies: \textit{none}, \textit{hourly}, \textit{random}, and \textit{calm-only}.
The \textit{none} condition provided no prompts beyond an end-of-day reflection.
The \textit{hourly} condition delivers regular, neutrally framed reflection prompts every hour during waking hours.
The parents' responses to surveys during this period are used to train our machine-learning-based model to enable the subsequent \textit{calm-only} condition.
The \textit{random} and \textit{calm-only} conditions included a lightweight positive reinforcement intervention: prompts encouraged parents to say something supportive to their child, either at randomized times or when low activity suggested a calm moment (See~\figref{fig:system-overview}).
The within-subjects design maximizes statistical power despite the small sample size and accommodates substantial individual differences in parent-child dynamics and children's behavioral patterns, making it well-suited for our probe objectives.

\subsection{Procedure}

The study procedure is presented in~\figref{fig:study-overview}.
First, we conducted in-person onboarding, during which participants provided informed consent, received an Apple Watch, were introduced to the watch interface, installed the phone app on their device, and completed a pre-study survey.
After that, all participants experience conditions in the same order for the first two weeks (none → hourly), followed by counterbalanced conditions (random and calm-only) in weeks 3-4.
The fixed order for weeks 1-2 is necessary because baseline data must precede intervention, and sufficient labeled data must be collected before implementing personalized models.
The counterbalanced weeks 3-4 control for ordering effects while ensuring all families experience both novel timing approaches.

\paragraph{Week 1: None Condition}
Children wore the smartwatch, and the parent completed only end-of-day surveys without receiving intraday notifications.
This established natural behavioral patterns and baseline measures of parental awareness, relationship quality, and parent-child dynamics without intervention.
This week also allowed families to acclimate to the smartwatch and daily survey routine. 

\paragraph{Week 2: Fixed Hourly Reflection Prompts}
Parents received hourly reflection prompts between 8 AM and 8 PM, in addition to the daily evening survey (as shown in~\figref{fig:system-overview}).
This phase supported two goals: (1) to increase parent awareness about their children's actual activity levels through regular, neutrally framed moments of reflection throughout the day; and (2) to collect parents' perception data for tailoring our motion-based model.

\paragraph{Week 3-4: Counterbalanced Reinforcement Prompts}

Participants were randomly assigned to receive two reinforcement prompting strategies in counterbalanced order using block randomization: random and calm-only.
For the random condition, participants received five positively framed notifications per day, delivered at unpredictable times between 8 AM and 8 PM.
This strategy introduced positively framed prompts in a variable rhythm to explore how families respond to unpredictability in reflective moments.
On the other hand, in the calm-only condition, participants received positively framed prompts whenever our motion-sensing pipeline detected calm moments, using a model personalized with Week 2 data.
This strategy explored how families engage with prompts aligned with inferred calm moments, where both the presence and absence of a notification could carry meaning.

\subsection{Survey and Data Collection}
During the course of the study, we collected both quantitative behavioral data (response rates, timing patterns) and qualitative user experience data (perceptions, usability feedback) through multiple survey instruments administered at different temporal scales, as presented in~\figref{fig:study-overview}.
 
\paragraph{Pre-study Survey:} This is administered during onboarding to establish baseline characteristics and allow participant segmentation. 

\paragraph{Intraday Surveys (Weeks 2-4 only):} During weeks 2 to 4, we also administered a brief intraday survey. These contained 2 items: (1) child activity level using a 5-point scale from ``Very Calm: Mostly sitting quietly'' to ``Very Active: Constant movement,'' and (2) open-text description of child's recent activities.

\paragraph{End-of-Day Surveys (All weeks):}
At 8:00 PM daily, participants were sent a survey containing three items: 
(1) medication adherence (yes/no), 
(2) parent-child communication effectiveness (``How effectively did you think you communicated with your child today?'' rated on a 5-point Likert scale from ``Not effective at all'' to ``Extremely effective''), 
and (3) a free-form reflection textbox for the day.

\paragraph{End-of-Week Surveys (All weeks):} At the end of every week, participants were sent an end-of-week survey. The weekly assessments contained 12 items across five domains: (1) open-text child behavior description, (2) parental self-efficacy/overwhelm using items adapted from established scales (\eg ``My interactions with my child this week make me feel effective and confident as a parent'' and ``I felt overwhelmed by the responsibility of being a parent this week''), (3) parent-child relationship quality (items on closeness and positive reinforcement), (4) technology evaluation including notification awareness, connection quality, and manageability (Weeks 2-4 only), and (5) open-text reflections on learning and behavioral changes. All Likert items use 5-point scales from ``Strongly Disagree'' to ``Strongly Agree.'' 

\paragraph{Exit Interview:} At the end of the four weeks, we conducted a semi-structured interview covering overall experience, comparative evaluation of calm-only versus random notification conditions, impact on daily life and parent-child communication, system feedback, and perceived changes in family interactions. We analyzed the parents' responses in the exit interviews using open coding~\cite{StrCor90}.
Two authors first independently coded the emerging themes and then discussed them until reaching agreement.

\section{Results}
\label{sec:results}

Twelve out of the sixteen parent-child dyads completed the study, while four dropped out midway, citing burden from study participation and external circumstances.
Most dropouts occurred during the hourly condition (D2, D3, D4) and one happened early in Week 3 (D1), with all reporting high baseline stress levels. Three of these participants withdrew officially by emailing the study team, while one (D4) stopped responding to the surveys and our follow-up emails. Prior to dropout, these participants showed similar engagement patterns to completers.
We discuss this result further in~\secref{sec:results-stress}.

All participants who completed the study responded to the end-of-week surveys, with the exception that two parents missed the final questionnaire before it expired, still enabling condition-wise comparisons.
The response ratios to the end-of-day and intraday surveys are summarized in~\tabref{tab:result_response_rate}.
These results indicate that the parents who completed the study engaged consistently across the four-week period.

In the rest of this section, we describe the insights we gained from our user study using both quantitative and qualitative methods to answer the research questions outlined in~\secref{sec:study-rqs}.

\subsection{Motion Sensing can Approximate Parental Perceptions of Calm Moments, while Alignment Varies across Families (RQ1)}
\label{sec:result-rq1}

We first evaluated the goodness of fit of the regression model.
We randomly split the dataset of acceleration energy and response score pairs in the hourly condition into training and test sets (8:2).
A single global regression model achieved $R^2 = 0.25$.
A personalized regression model trained separately for each participant, while preserving the same overall train–test split, achieved $R^2 = 0.44$, indicating a moderate level of predictive performance.

We then evaluated the proportion of moments in which parents perceived their child as calm (\ie response $< 3$) at the time they received a notification.
While the personalized regression model achieved a moderate $R^2$ of 0.44 (in the hourly condition dataset), the perceived calm ratio was 78.0\% in the calm-only condition.
This implies that parents indeed received notifications during calmer moments and that thresholding the outputs to derive notifications highlights the practical utility of the regression model.
Additionally, the perceived calm ratio was 38.7\% for the random condition, which closely matches the overall label balance observed in the hourly condition data, where 42.9\% of responded instances were perceived as calm.

These results suggest that while the acceleration energy model can capture parents' perceived calm moments, the alignment varies across families.
This family-specific variability is consistent with prior work on ADHD, which has documented discrepancies in how caregivers perceive and rate children's ADHD and hyperactivity symptoms, including systematic differences between mothers and fathers and only low-to-moderate agreement between parents and teachers~\cite{Chen2017Relationship,Sollie2012Comparison}.
These findings underscore that parents' perceptions of their child's hyperactivity and calmness are shaped by family-level factors, and motivate our use of personalized models that adapt to each family's data.

Lastly, while our findings demonstrate the potential of the calm-detection model, it is challenging to draw rigorous conclusions regarding its accuracy from this in vivo deployment.
Ground-truth labels (parents' responses) were only collected at the moments when parents received notifications \textit{and} responded, limiting our ability to compute metrics such as recall or F1-score.
A more controlled study with continuous ground-truth annotation would be necessary to fully evaluate these metrics.

\begin{table}[t]
\centering
\small
    \caption{Response rate (\%) across the completed 12 parents by condition}
    \label{tab:result_response_rate}
    \begin{tabular}{r|c|c|c|c}
    \toprule
     & None & Hourly & Random & Calm-Only \\
    \midrule
    Intraday & N/A & 59.6 $\pm$ 16.6 & 59.0 $\pm$ 17.7 & 64.8 $\pm$ 20.0 \\
    End-of-Day & 78.4 $\pm$ 24.5 & 74.1 $\pm$ 28.6 & 62.5 $\pm$ 19.3 & 74.3 $\pm$ 18.1 \\
    \bottomrule
    \end{tabular}
    \Description{Table displaying the response rate in percentages across all 12 completed parents by condition. There are two rows for each notification type (Intraday and End-of-Day), and four columns with the conditions (None, Hourly, Random, and Calm-Only).}
\end{table}

\subsection{Flexible Adoption of \system{}'s Features: Parents as Active Designers of Their Own Interventions (RQ2+RQ3)}
\label{sec:result-rq2}

\begin{figure*}[t]
    \centering
    \begin{subfigure}[b]{\textwidth}
        \centering
        \includegraphics[width=\textwidth]{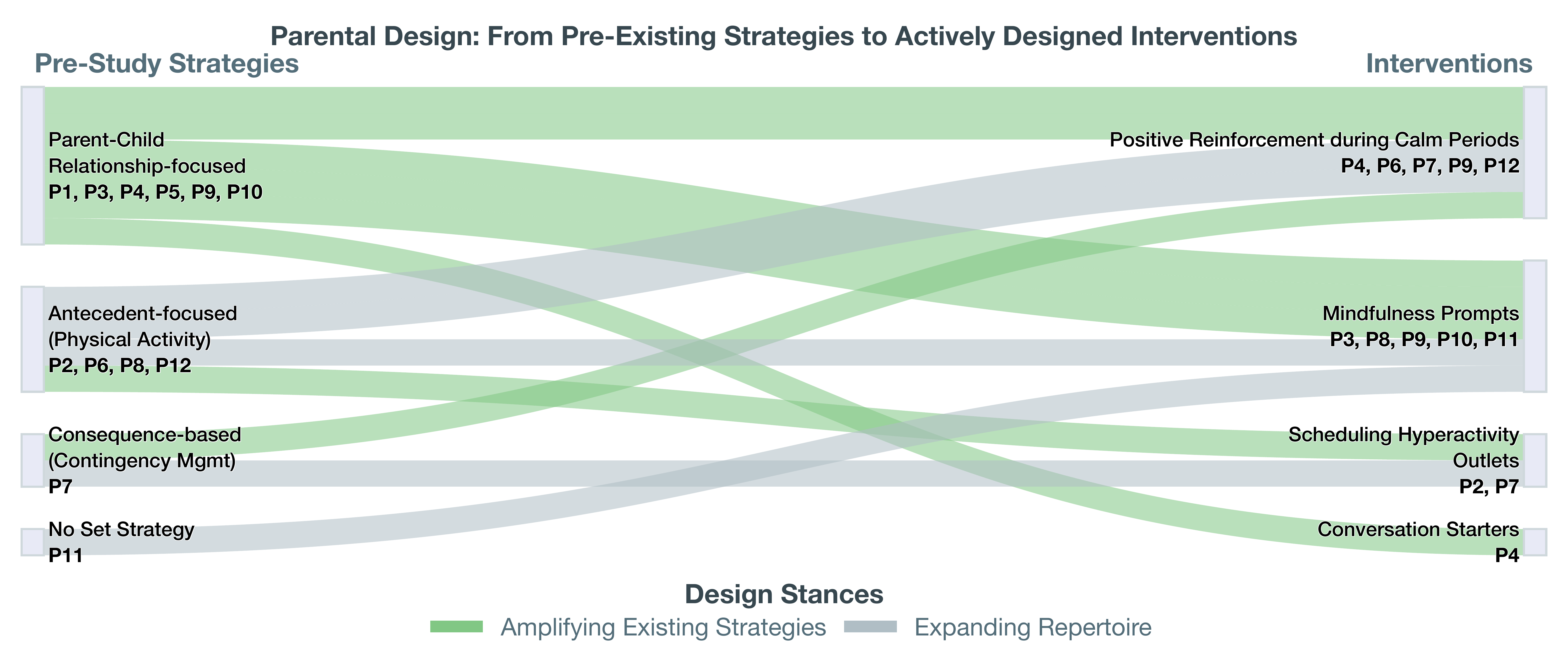}
    \end{subfigure}
    \caption{Parents with diverse pre-existing management strategies (left) appropriated CalmReminder features to design distinct interventions (right) through two mechanisms: amplifying existing practices (green flows) or expanding into new management domains (gray flows). Most relationship-focused parents (P3, P4, P9, P10) amplified their relational approaches, while three of four antecedent-focused parents (P6, P8, P12) expanded from physical to psychological strategies. P1 and P5 did not adopt features during the study period.}
    
    \Description{Sankey diagram illustrating how parents transitioned from their pre-study strategies to actively designed interventions during the study. The left side shows four categories of Pre-Study Strategies: (1) Parent-Child Relationship-focused (P1, P3, P4, P5, P9, P10), (2) Antecedent-focused (Physical Activity) (P2, P6, P8, P12), (3) Consequence-based (Contingency Management) (P7), and (4) No Set Strategy (P11). The right side shows four Intervention Strategies co-designed or chosen during the study: Positive Reinforcement during Calm Periods (P4, P6, P7, P9, P12), Mindfulness Prompts (P3, P8, P9, P10, P11), Scheduling Hyperactivity Outlets (P2, P7), and Conversation Starters (P4). Green paths represent the Amplification of existing strategies, while gray paths denote Expanding the parental repertoire with new strategies. For example, P4, who initially used a relationship-focused approach, expanded their strategies to include Conversation Starters and Positive Reinforcement. P11, who had no pre-set strategy, adopted Mindfulness Prompts, illustrating the introduction of new design tools during the intervention process.}

    \label{fig:adoption-sankey}
\end{figure*}

\system{} provided multiple new features - calm detection, varied intraday notification timings across study weeks, and positive reinforcement reminders as a way to encourage parents to give positive reinforcement to their children during calm moments. Outside of the positive reinforcement prompts sent, we did not prescribe a single ``correct'' way to use these features. This partially open-ended design choice revealed a striking pattern: most parents did not passively consume the prompts or accept/reject the system as a whole. Rather, they actively functioned as \textit{designers} of their own interventions.
Based on their family contexts, parents made design-level decisions about \textit{which} features to use, \textit{what} problems to address, \textit{how} to involve their child, and \textit{when} to intervene.
We did not notice any substantial correlations between the ADHD hyperactivity scores or parental stress and the kind of intervention they actively designed for parents who adopted the system.
Instead, these design patterns were shaped primarily by compatibility with parents' existing philosophies.
We observed two distinct design stances:

\begin{itemize}
    \item \textbf{Amplifying Existing Strategies}: Several parents (P2, P3, P4, P7, P9, P10) used \system{} to reinforce practices they already employed. 
    For instance, P2 repurposed calm-moment notifications as reminders to provide physical activity outlets, while P4 used positive reinforcement prompts to increase the frequency of her existing relationship-focused conversations. Similarly, P9 used the notifications to be more mindful of her overall parenting and to use more compliments in general, amplifying her relationship-focused strategy.
    These parents selectively engaged with system features that supported their established approaches.
    \item \textbf{Expanding Repertoires}: Parents (P6, P7, P8, P11, P12) used \system{} to experiment with novel practices in areas where they had not previously formed deliberate strategies.
    P8, for example, transformed notifications into touchpoints for parental self-reflection: a practice they had not previously engaged in. P6 tested calm-moment praise, observed its positive effects, and incorporated it into her parenting repertoire.
    These strategies represented additions to parents' existing repertoires rather than wholesale replacements.
\end{itemize}

These design patterns and interventions emerged naturally during deployment rather than through formal participatory sessions. Though not explicitly designed as a self-experimentation platform, \system{}'s open-ended design enabled parents to experiment with, evaluate, and selectively adopt interventions - extending self-experimentation frameworks~\cite{karkar2017tummytrials, daskalova2021self} from individual to family contexts.

\figref{fig:adoption-sankey} summarizes how families appropriated different system features across these intervention patterns. All relationship-focused parents who adopted some features (P3, P4, P9, P10) did so in a way that amplified their existing approaches, potentially because \system{}'s features matched their relational philosophy, while three of the four antecedent-focused parents (P6, P8, P12) expanded from physical activity management to include new psychological tools like mindfulness. This suggests parents amplified when system features aligned with their existing practices, but were often open to expanding their repertoires when features introduced fundamentally new management approaches. 
Below, we further analyze how parents actively shaped interventions by appropriating \system{}'s features in ways that aligned with their parenting philosophies and everyday contexts. 

\subsubsection{Positive Reinforcement during Calm Moments (P4, P6, P7, P9, P12)}
\label{sec:result-rq2-positive}

Five parents used \system{} \textit{as designed} to provide positive reinforcement during calm moments. These parents consistently engaged with intraday notifications to receive behavioral check-in prompts and reinforcement timing cues, and utilized positive reinforcement reminders as intended for calm-moment praise.
Yet even this `as designed' usage reflected active design decisions. Parents didn't simply execute a prescribed protocol; rather, they made interpretive choices about what positive reinforcement during calm-moment notifications meant within their family dynamics and crafted individualized strategies for implementing reinforcement that aligned with their existing parenting philosophies and communication styles.

P4, P7, and P9 amplified relationship-focused strategies, and used positive reinforcement as a way to enhance positive interactions with their child. P9, for instance, found the reminders helpful because, as a busy parent, she struggled to notice calm moments, and the prompts helped her foster more positive interactions.
On the other hand, P7 selectively interpreted features to fit her consequences-based management approach, valuing notifications during calm moments but also during hyperactive moments to \emph{``let [her child] know he's either...doing a good job, or...being hyperactive times 10.''} 
In contrast, for P6 and P12, the intervention proved to be a helpful segue into positive parenting strategies. P6 reached this intervention through active exploration, experimenting with calm-moment praise, observing it made her child \emph{``feel good about it''} and integrating it into her repertoire.

Similarly, non-adoption of this intervention often represented active evaluation rather than passive disengagement. Parents rejected the prescribed intervention through two pathways: \textit{empirical rejection}, where they experimented before discontinuing (\eg P2, whose child found compliments awkward and inconsistent with her active-play philosophy), and \textit{conceptual rejection}, where they declined without trial based on existing beliefs (\eg P1, who deemed her pre-existing strategies sufficient).
These pathways reveal varying levels of experimentation parents were willing to undertake when evaluating system-suggested practices, which also depended on their pre-existing philosophies and confidence in those.

\begin{insightbox}
\textbf{Parental Design Insight:} Active design occurs even when parents use features as intended; they still decide how those features fit their contexts and philosophies. This means that apparent ``correct'' use often masks meaningful variation in how families shape interventions. Systems should therefore make these interpretive choices visible — for example, by offering lightweight configuration options or reflective onboarding that helps families clarify how features align with their goals.
\end{insightbox}

Beyond this primary intervention, parents actively repurposed system features to design alternative support strategies. Interestingly, these emergent interventions were not always implementations or appropriations of existing practices - they often reflected parents' aspirations to develop new capabilities and improve their approaches. Parents not mentioned for each intervention did not discover those particular interventions.

\subsubsection{Notifications as Mindfulness Prompts (P3, P8, P9, P10, P11)}
\label{sec:result-rq2-mindfulness}
Five parents appropriated the system's notification structure (\ie intraday notification and positive reinforcement) to cultivate heightened awareness of their parenting practices, as well as their child's behavioral patterns. Rather than responding to prompts in the moment, these parents used notifications as grounding touchpoints for self-reflection, subsequently adjusting their interactions with their children through increased compliments or calmer communication, or paying closer attention to their child's behavior.
For instance, P9 mentioned,
\begin{quote}
    \emph{``
    I think those positive reminders... I think with anything, were grounding to be like, have I given those to him today?''} and continued to mention \emph{``If I notice, like, oh, I've been screaming all day because, like. the day, like, it helped kind of ground me to, like, be more mindful of that and use more compliments. ''} (P9)
\end{quote}
This comment suggests that the participant noticed an increase in the amount of positive reinforcement to her child by the designed reminder. 
P8 also thought \system{} helped them reflect on their general parenting strategies and incorporate more positive parenting:
\begin{quote}
    \emph{``I think it made me more mindful of... You know, his... his... The way he is, and him as a person, and being able to, sort of, like, hold myself back from maybe, like, if I wasn't, in a good mood or something, I'd be like, okay, well, this isn't really his fault, like, you know, I could, speak to him more, like, calmly''} (P8)
\end{quote}

Parents also used the notifications to foster understanding and compassion for their children. P10 noted that receiving notifications helped her feel a sense of compassion and understanding as her child works through tough moments:
\begin{quote}
 \textit{``And it's just, I think, given me a different appreciation for, like, how difficult it really is for her to function like this....So, I think it's made me probably more compassionate and just, like. Okay, it's not her, it's her brain, her brain's working against her, just, like, calm down, it's fine.''} (P10)
\end{quote}

This result suggests that parents with existing positive reinforcement practices or even those who did not agree with giving positive reinforcement during calm moments could extract additional value from \system{} by expanding its scope beyond hyperactivity management towards parental reflection.

\begin{insightbox}
\textbf{Parental Design Insight:} Parents may repurpose child-focused prompts as parent-focused mindfulness cues, using notifications for self-reflection rather than following their instructional framing. This shows that parents design for dual purposes — immediate support for the child and their own longer-term capability building. Systems should therefore offer features flexible enough to support both moment-to-moment guidance and ongoing parental development.
\end{insightbox}

\subsubsection{Reminder for Scheduling Hyperactivity Outlets (P2, P7)}
\label{sec:result-rq2-outlets}

Two parents repurposed calm detection as an activity awareness tool.
Often, an existing family strategy regularly schedules activities to provide productive outlets for their children~\cite{DEKKERS2022478,Leijten2019}.
In such cases, \system{}'s intraday notifications helped parents be more aware of when their child might need these outlets.
For instance, P2, whose existing management strategy focused on finding activity outlets, found \system{} to be useful in increasing her awareness of her child's activity level.
She used the notifications as a way to identify when her son had not had enough opportunities for physical activity and to be more cognizant of his needs.
She specifically noted that her child required more physical activity than she realized, and was able to utilize the notifications to bring awareness to this need: 
\begin{quote}
    \emph{``I was looking at sort of his activity and lack thereof and so I think that helped me to think about [how] his activity diet needs to be more so I think that was helpful.''} (P2)   
\end{quote}
Similarly, appreciating how \system{}'s notifications helped her notice her child's hyperactivity levels more, P7, who focused on contingency management, adopted the notifications as a reminder to schedule more outlets mentioned - something she had not intentionally tried to focus on before.
\begin{quote}
    \emph{``I just think it's good to know, like, when he... when he is being hyperactive. I mean, I think... because then you can use, you know, different type of coping tools. I don't know what the word to use that, to help him get out of that.''}  (P7)
\end{quote}
These accounts suggest future possibilities of our real-time sensing approach beyond calm detection to enable parents to select strategy-focused intervention timings. 
While such designs must be implemented with care regarding children's privacy, we discuss concrete implications in~\secref{sec:disc-jitai}.

\begin{insightbox}
\textbf{Parental Design Insight:} Parents may repurpose calm detection as a general activity-awareness signal, creating their own interpretations of sensor-driven cues. Systems should therefore expose flexible, privacy-aware activity primitives that families can map onto diverse management strategies, while calm-moment interventions remain effective and clinically recommended.
\end{insightbox}

\subsubsection{Trigger for Conversations (P4)}
\label{sec:result-rq2-conversation}

One parent repurposed \system{}'s intraday notification to serve as a catalyst for initiating conversations with her child.
Her child had recently been diagnosed with ADHD, and their psychologist had recommended discussing the reasons behind different behaviors, and the app allowed her to increase the frequency of the conversations.
P4 used notifications as prompts to draw attention to these behaviors and to frame them within the broader context of ADHD:
\begin{quote}
\emph{
    ``He's newly diagnosed, so I think that it [\system] was really helpful to kind of call attention to his behaviors. His psychologist, as well, told us, as we moved forward, kind of calling attention to, `Hey. You behave this way because your brain works a little bit differently, you have attention deficit disorder, and kind of calling attention to the different things that he's doing, and I think that the application kind of helped bring that a little more to the forefront.''
    } (P4)
\end{quote}
This illustrates how notifications can extend beyond behavioral regulation to support parent-child dialogue and shared understanding of ADHD.
Such use aligns with therapeutic recommendations and suggests that future systems could be designed not only to guide parent interventions, but also to scaffold family communication and the child's self-awareness.

\begin{insightbox}
\textbf{Parental Design Insight:} 
Parents may also design dyadic interventions by intentionally making system features visible to their children. Systems should therefore consider how features can be appropriated for parent-child collaborative sensemaking around behavior-based interventions, rather than assuming parent-only use.
\end{insightbox}

\subsubsection{System Rejection (P1, P5)}
\label{sec:result-rq2-rejection}

Two parents reported that they did not gain additional insights from \system{}.
Both parents had pre-existing strategies rooted in a parent-child relationship focus, and claimed that their existing strategies already incorporated enough positive reinforcement. They felt that \system{} did not add any value for them. For example, P5 noted that reminders to positively reinforce her child would come at times when she would have already done the reinforcement. While this reconfirms that our motion sensing was aligned with parental perception, it also highlights that the intervention was not adding sufficient value for parents who already had consistent strategies in place.

\begin{insightbox}
\textbf{Parental Design Insight:} 
For parents with strong, consistent reinforcement strategies, \system{} contributes little additional value beyond what they already practice.
\end{insightbox}

\begin{figure*}[t]
    \centering
    \begin{subfigure}[b]{\linewidth}
        \centering
        \includegraphics[width=\linewidth]{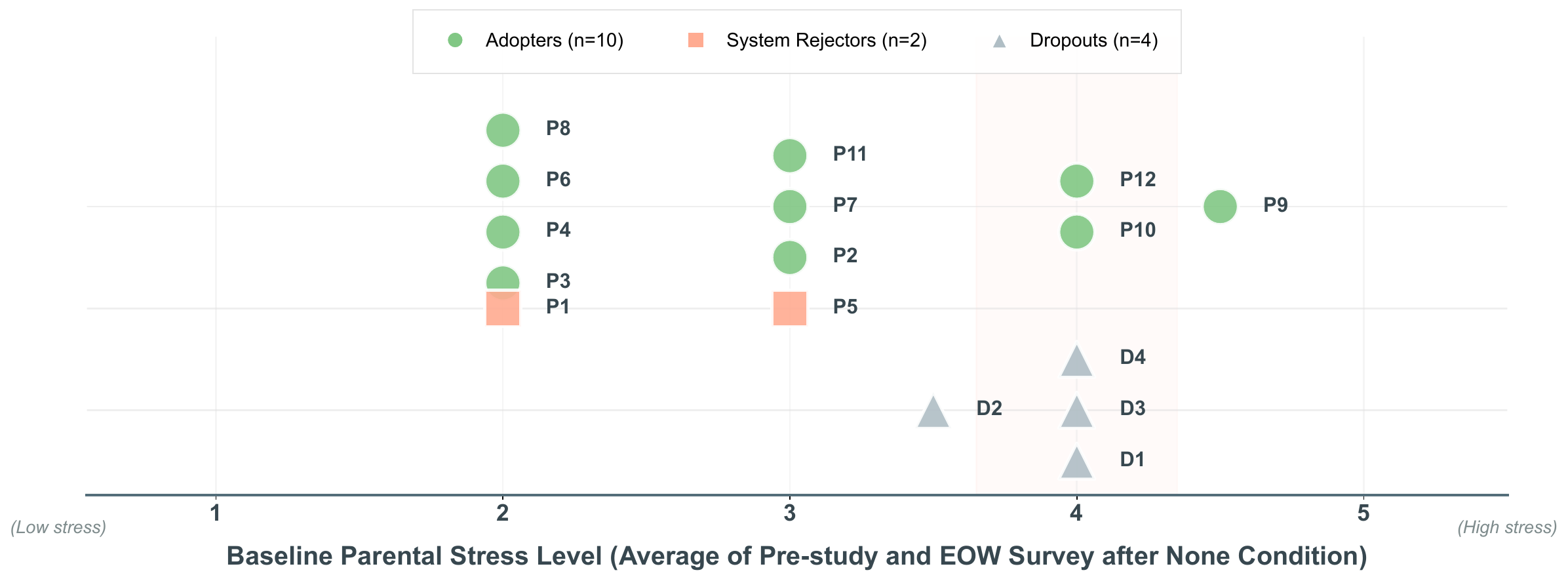}
    \end{subfigure}
    \caption{Baseline self-reported parental stress as a gating factor for system adoption. Distribution of participants across baseline stress levels (1=low, 5=high) by adoption outcome. Adopters (n=10, green circles) are distributed across all stress levels, and most have lower baseline stress scores. Dropouts who withdrew mid-study (n=4, gray triangles) - cluster exclusively at high stress levels ($>3$). This pattern suggests that extreme baseline stress precludes initial engagement, rather than as a continuous predictor of appropriation patterns among those who do engage.}
    \Description{A scatter plot showing baseline parental stress level (x-axis, ranging from 1 to 4) versus adoption outcomes (distinguished by color and shape). The plot uses three distinct markers: green circles for adopters, orange squares for system rejectors, and gray triangles for dropouts. At stress level 1, there is one green circle labeled P6. At stress level 2, there are three green circles labeled P2, P3, and P4, arranged vertically. At stress level 3, there is one green circle labeled P8 and one gray triangle labeled D2. At stress level 4, there are two green circles labeled P7 and P9 at the top, two orange squares labeled P1 and P5 in the middle, and three gray triangles labeled D1, D3, and D4 at the bottom. Two key patterns are annotated: an arrow pointing to the stress level 4 cluster notes "All non-adopters (rejectors + dropouts) at high stress," while an arrow pointing to the adopters notes "Adopters spread across all levels." A subtle shaded region highlights the stress level 4 area where non-adopters cluster. The visualization demonstrates that all six non-adopters (two rejectors and four dropouts) had baseline stress levels of 3 or 4, while the seven adopters were distributed across the full stress range from 1 to 4.}
    \label{fig:result_stress_level}
\end{figure*}

\subsection{High Baseline Stress May Hinder Initial Adoption (RQ2)}
\label{sec:results-stress}
We calculated the baseline parental stress levels by averaging out the pre-study and end-of-week survey responses to the ``I felt overwhelmed by the responsibility of being a parent this week'' question.
The dropout group: 3 of 4 (D1, D3, and D4) dropout participants reported baseline stress of 4, while the remaining participant (D2) reported 3 (See~\figref{fig:result_stress_level}).
While we do not draw statistical conclusions, the trend suggests that high baseline stress may shape the initial adoption process. 4/7 participants who had a high baseline stress ($>$ 3) did not adopt the intervention, while 7/9 parents with lower scores adopted at least some features of \system{} and none dropped out.
For instance, D1, who dropped out during the start of the third week, did not fill out the EOW survey for the hourly condition but in the exit interview echoed how \textit{``the notifications were more annoying and stressful than helpful''}, adding that \textit{``it wasn't helping, it was frustrating me, and distracting to him.''}
The rest of the dropped-out participants refused to interview and dropped out before they could fill out the EOW survey for the hourly condition.

At the same time, we also found that parents with similar stress scores often adopted different strategies, and conversely, parents using the system in similar ways exhibited varying stress levels.
For instance, two parents who repurposed the system for mindfulness had divergent stress levels (2 for P3 and 4.5 for P9).
This pattern suggests that while extreme stress may preclude initial adoption, appropriation strategies among adopters are shaped primarily by existing practices and parenting philosophies rather than stress levels alone.

\begin{insightbox}
\textbf{Parental Design Insight:} High baseline stress functioned as a barrier to active design: parents reporting extreme stress often rejected the system or dropped out, while parents who had lower baseline stress often adopted the system in some form. Active design and high-frequency interaction require resources (time, cognitive bandwidth, willingness to experiment) that overwhelmed families cannot spare. Systems should provide minimal-effort baseline functionality for high-stress users while enabling deeper customization for those with the capacity to engage.
\end{insightbox}

\subsection{Varied Impressions of \system{}'s Prompting Strategies across Families (RQ3)}
\label{sec:result-rq3}

\begin{figure*}[t]
    \centering
    \includegraphics[width=0.9\linewidth]{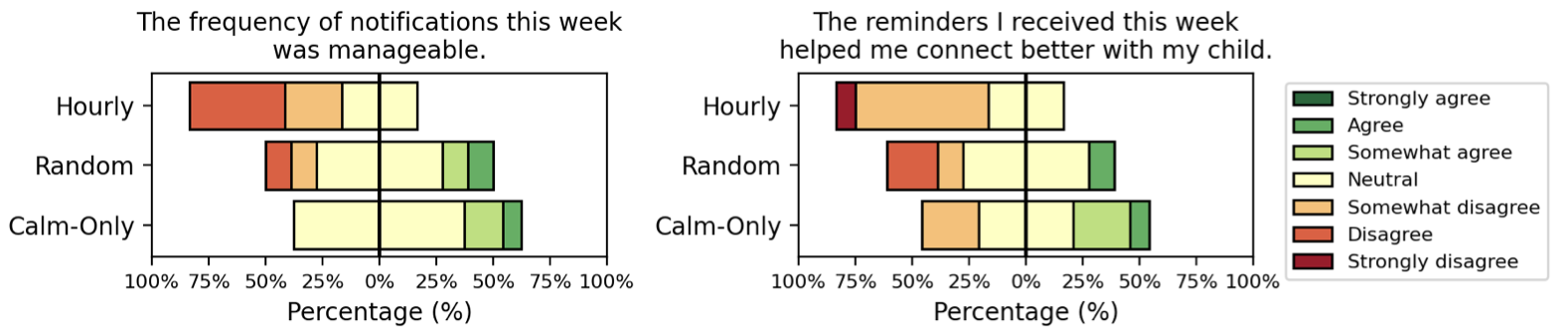}
    \caption{Participants' end-of-week survey responses on \system{}'s intraday notifications, rated on a 1–7 Likert scale.}
    \Description{Two diverging stacked bar charts that describe the participant's end of week survey responses on \system{}'s intraday reminders. Both diverging stacked bar chart shows the conditions (Hourly, Random and Calm-Only) on the y-axis, and percentage on the x axis. The bars indicate the distribution of responses based on a 1-7 Likert scale rating. The topmost diverging stacked bar chart displays parent responses for the question "The frequency of notifications this week was manageable." The bottom bar chart shows the parent response for the question "The reminders I received this week helped me connect better with my child."}  
    \label{fig:result_likert}
\end{figure*}

Finally, we analyzed the parents' responses to the end-of-week surveys together with their responses during the exit interviews to explore how different notification strategies affected their perception and adoption of \system{}.
Responses related to parental self-efficacy and parent-child relationships showed little variation across conditions, suggesting that these aspects are influenced more by the broader context of daily life than by \system{}'s interventions.
In contrast, although the limited number of participants prevents us from drawing statistical conclusions, we observed emerging trends (\figref{fig:result_likert}), which we elaborate on in the following subsections.

Note that the hourly condition sent 12 notifications a day and had to precede the calm-only condition as the responses enabled model training, which is our study requirement.
Given these differences, the findings below are presented as exploratory descriptions of parents' experiences across these conditions rather than a controlled comparison.

\subsubsection{Mixed but Meaningful Impacts of Hourly Prompts}

Survey responses at the end of the hourly week indicated that most parents found the hourly notifications difficult to manage, and did not feel they helped them connect better with their child.
Exit interviews conducted after all conditions were completed, however, revealed more nuanced, family-specific perceptions as parents reflected on their full experience.
For some parents, the constant interruptions were burdensome, adding stress to already full days and even contributing to one participant's dropout, mentioning,
\begin{quote}
    \emph{``With busy lives, it's... annoying...I, like, couldn't even say anything because I had no idea what was going on with them. Got it. And for any working parent or for a child in school, it's just not plausible to know what's going on and to answer those surveys, so....''} (a dropped participant)
\end{quote}
However, interestingly, multiple parents (P2, P4, P7), who valued structured reflection, appreciated the frequent hourly prompts, even though they were working, and it was often sent when the children were at school. 
For these families, frequent prompts functioned as reliable cues that sustained awareness and created repeated opportunities for reflection and noticing their children's activities.
P2 noted that the frequent notifications helped her become more aware of her child's activities: 
\begin{quote}
    \emph{``It prompted me to be intentional with my communication and proactive in setting up a good dynamic between the two of us. And then, we had a really great week, and I think that the notifications made that happen.''} (P2)
\end{quote}
Similarly, P4 and P7 noted how they preferred hourly notifications as this helped them reflect and acted as reminders: for example, 
\begin{quote}
    \emph{``I mean, granted, I did miss some notifications and stuff like that, depending on if I was working or whatever, but, overall, I felt like it wasn't terrible to keep up with or pay attention to. I'm always glad... I also am ADHD, so I... I appreciate reminders, and, like, any kind of checklist for me, so, like, it'll bring stuff to my mind as well.''} (P4)
\end{quote}
These nuanced perceptions suggest that high-frequency notifications are not universally experienced as fatigue-inducing; rather, their value depends on parents' orientation toward reflection and their existing management practices, while sporadic notifications were generally perceived as more manageable and less intrusive.

\subsubsection{Positive but Varied Effects of Random and Calm-Only Prompts}

Most participants were neutral/above for the manageability of the notifications in both the random and calm-only notifications. Parents' responses to whether the notifications helped them connect better with their child showed variation during these conditions: some found value, while others remained neutral/disagreed.
This effect may reflect the value of positive reinforcement prompts in these conditions for at least some parents, and the variation in the designed interventions discussed in~\secref{sec:result-rq2}.
Notably, most parents did not perceive a clear difference between the calm-only and random strategies, while some parents adopted the positive reinforcement during the calm moments (\secref{sec:result-rq2-positive}).
One possible explanation is that notifications delivered at \textit{non-calm} moments were still seen as useful—for example, as mindfulness prompts, opportunities to schedule outlets for hyperactivity, or conversation starters as discussed above. 
Although the limited number of participants prevents us from drawing statistically conclusive claims, the observed trend reconfirms the differing patterns of adoption, where some parents appreciated calm-only prompts while others found other strategies similarly useful or unobtrusive.
This suggests that a one-size-fits-all approach is unlikely to be effective.
Instead, systems should support flexible, family-specific ways of engaging with prompts, a design direction we further elaborate in~\secref{sec:disc-jitai}.

\section{Discussion}
\label{sec:disc}
Our four-week user study revealed nuanced implications for the design of systems that deliver prompts to support parental reflection and engagement around hyperactivity symptoms in children with ADHD.
As our results demonstrated in~\secref{sec:result-rq2}, the adoption pattern of \system{}'s different features is highly influenced by the family's existing parenting strategy and values. 
For instance, a majority of parents actively designed interventions that amplified rather than replaced pre-existing strategies.
These findings align with prior work on co-regulation technologies~\cite{DBLP:conf/chi/SilvaCMBBBESLH23, DBLP:conf/chi/SonneMMOG16} and family informatics~\cite{DBLP:conf/chi/KientzAA09, DBLP:conf/chi/HinikerSSSLK15}, which highlight that technologies are most effective when they amplify rather than overwrite existing routines and values.
In line with the person-based approach~\cite{Yardley2015PBA, Yardley2015PBAResult}, future systems should provide \textit{configurable features} that families can tailor to their management strategies and capacity for experimentation.
In practice, our findings suggest offering families control over the role and intensity of the prompts; whether they function as positive reinforcement, mindfulness, conversation starters, or activity scheduling.
This recommendation underscores the need to personalize \emph{when} and \emph{how} \system{}'s prompts are surfaced within families' daily routines, which we further elaborate on in what follows.

\subsection{Behavior-Based Prompting for Just-in-Time Adaptive Interventions in Family Settings}
\label{sec:disc-jitai}

Prior HCI work and the broader literature on Just-in-Time Adaptive Interventions (JITAIs) have consistently shown that timing shapes both engagement and intervention effectiveness~\cite{Morrison2017, Bidargaddi2018, NguyenTran2025, NahumShani2017JITAIs}.
Existing systems operationalize ``good timing'' in different ways, for example, through scheduled prompts that ensure consistency~\cite{Fry2009}, context-aware systems that detect user availability~\cite{8115839}, and adaptive algorithms that learn personalized patterns of receptivity~\cite{Wang2021}.
Our work contributes a complementary perspective on JITAI timing: \emph{the child's behavioral state itself} can serve as a key timing signal for parents.
By detecting children's low-hyperactivity periods, \system{} surfaces moments when parents are more receptive, less reactive, and better able to engage in constructive interactions—conditions that behavior-change theories describe as windows of opportunity rather than moments of crisis~\cite{NahumShani2017JITAIs}.

Empirically, we observed flexible adoption of the calm-moment-based intervention across parenting styles, as discussed in~\secref{sec:result-rq2}, including how parents used prompts to notice, reinterpret, and reflect on their child's behavior.
Here, we see such behavior-based prompting as a novel timing strategy for future adaptive JITAIs in family settings.
First, our findings on the parental varied appreciation of different notification strategies (\secref{sec:result-rq3}) point to the value of offering multiple, configurable timing mechanisms that families can select.
Second, while our study intentionally focused on calm moments per clinical guidance, exploring prompts during escalating moments presents an opportunity for future systems—potentially complementing calm-moment prompts by supporting preparation, as suggested in~\secref{sec:result-rq2-outlets}.
Importantly, our work shows that such real-time motion sensing is feasible with an off-the-shelf device, and that the alignment between motion-derived energy signals and parental perception varies across families (\secref{sec:result-rq1}).
To facilitate further exploration of this space, we will release our system code (watch and server) as open source.

\subsection{Design Implications for Dyadic Reflection and Reinforcement}
\label{sec:disc-dyadic}

Furthermore, our findings suggest that the value of just-in-time support in family settings lies not only in \emph{when} an intervention is delivered, but also in \emph{how} it shapes the relational processes surrounding parent-child interaction.
This relational perspective naturally extends beyond timing toward questions of shared reflection and co-regulation within the parent-child dyad.
The Theory of Dyadic Illness Management~\cite{Lyons2018Theory} emphasizes shared understanding, collaboration, and mutual support, highlighting the value of treating illness management as a relational process.
Although originally developed for adult–adult dyads, its core principles—shared appraisal of symptoms, coordinated management behaviors, and interdependent outcomes—closely align with how ADHD is handled within parent-child dynamics. Managing ADHD is inherently dyadic~\cite{Whalen2009, GERDES_HOZA_PELHAM_2003}: parents structure routines, scaffold emotional regulation, and coordinate treatment, while children gradually build self-awareness and agency. With developmentally appropriate adaptations, dyadic illness management theory offers a useful lens for designing technologies that cultivate collaborative reflection and reinforcement in families.

To further promote dyadic reflection, however, it is critical to actively involve children in ongoing management and communication.
For instance, we found participants may use \system{} to spark conversations with their child (\secref{sec:result-rq2-conversation}), indicating that intraday notifications can serve as natural entry points for lightweight, child-inclusive activities.
Therefore, future iterations could design prompts that not only inform parents but also scaffold brief shared activities, such as acknowledging a calm moment together, sharing a brief note of appreciation (\eg ``you're really focused right now''), or inviting children to check in about how they are feeling. These prompts could be adapted to children's developmental stages, sensory preferences, and family routines, ensuring that dyadic engagement feels supportive rather than intrusive.
By positioning calm-moment detection as an opportunity for joint sensemaking rather than unilateral monitoring, systems like \system{} can help families build shared awareness and strengthen collaborative regulation practices.
In doing so, even parents with strong existing reinforcement strategies (\secref{sec:result-rq2-rejection}) may find added value in the system as a relational artifact.

Dyadic theory can also inform how systems promote positive reinforcement, particularly for caregivers managing substantial and often overwhelming responsibilities. 
Overwhelmed caregivers frequently struggle to adopt novel technological support. This issue is also evident in our findings discussed in \secref{sec:results-stress}. Yet, prior studies have rarely addressed this dimension. Future systems could incorporate forms of positive reinforcement directed at parents, recognizing their ongoing effort, persistence, and resilience alongside child-focused guidance.
Such support could take the form of brief affirmations, reflective prompts that help parents notice their own successes, or feedback loops that highlight patterns of effective co-regulation. The opportunity to use notifications for self-reflection and mindfulness was evident in our findings (\secref{sec:result-rq2-mindfulness}).
By valuing the parent's experience as part of the dyadic process, systems may help reduce feelings of burden and increase willingness to sustain engagement. By embedding both dyadic reflection and positive reinforcement into smartwatch-based interventions, systems like \system{} can strengthen relational dynamics while supporting parental well-being and promoting children's developmental growth.

\subsection{Lightweight Configuration for (Already) Stressed Participants}
\label{sec:disc-model}

In addition to dyadic reflection, technical adjustments are needed to better support parents with high baseline stress.
Three of the four participants who dropped out did so during hourly-notification condition, (Week 2) citing the burden of frequent input. These patterns suggest that, for already overwhelmed caregivers, systems should minimize upfront demands and avoid relying on continuous user responses.

One approach is to begin with a lightweight configuration — for example, initializing with a global model trained on prior families, reducing notification frequency, or using simplified response mechanisms — and then gradually personalizing the model over time.
In addition, recent work demonstrates that richer reflections can effectively personalize activity trackers via natural language processing~\cite{DBLP:journals/imwut/KhayamiWKLCLCK25}, while parents' reflections on technology use for supporting children with ADHD are similarly crucial for sustained engagement~\cite{Kennedy2025From}.
Building on this, systems could incorporate brief end-of-day reflections or free-form feedback to refine calm-moment detection in a more sustainable manner.
Positioning reflection as both a support mechanism for caregivers and a channel for ongoing model adaptation enables family-centered sensing systems to achieve better alignment with parents' needs while maintaining accuracy without imposing additional burden.

\subsection{Smartwatches as Relational Artifacts in Family Life}
\label{sec:disc-children}

While \system{} was designed for parents, children were drawn to the smartwatch as active co-participants.
Their engagement—sometimes reflective and sometimes distracting—suggests that the device functioned as a situational relational artifact.
In moments when children interpreted the watch's cues as meaningful or responded emotionally, the smartwatch took on relational qualities despite not being intentionally designed as such.
Some parents observed that the smartwatch could disrupt their child's focus.
One dropped participant noted, \emph{``To keep on track ... but right now, being so little, it's more distracting than helpful.''}
Conversely, several families reported that the smartwatch sparked meaningful interactions: the child's curiosity about the device prompted conversations, encouraged reflection, and surfaced awareness around activity and behavior.
In this way, the smartwatch acted as a \textit{relational artifact}—a tool that shaped and mediated interaction between parent and child, even when the child was not directly using the app.
This aligns with prior HCI framing of health technologies as social artifacts that reconfigure relationships and shared attention within family contexts~\cite{turkle2006encounters}.
Smartwatches, by offering a tangible artifact, support joint engagement and narrative, similar to how storytelling devices operate as boundary objects that cross domains of interpretation.

Moreover, many parents highlighted how responsibility over the device became a moment for child growth and agency, which is aligned with the findings reported by Oyg{\"{u}}r~\etal~\cite{DBLP:journals/pacmhci/OygurEC20} that children often view managing a device as an opportunity to practice responsibility and develop personal agency.
P3 shared, \textit{``He was very responsible ... started his session with no reminders. He remembered to charge it every night. ... I was really impressed ... that was a good dry run.''}
In such cases, the smartwatch enables coordination, responsibility, and reflection between parent and child.

To sum, we recommend framing family-centered wearable interventions not only as sensing tools but as relational artifacts whose value lies in balancing dual roles: supporting parent-child engagement through the device while also respecting the child's freedom, agency, and everyday routines.
By embracing this dual nature, designers can better accommodate diverse family contexts and foster constructive dynamics without defaulting to prescriptive caregiving.

\section{Limitations and Opportunities for Future Work}

While our study successfully examined parental adoption of \system{} and yielded design implications, it also has several limitations.
First, the study involved a small number of participants; expanding the participant pool, especially including greater diversity in both parent and child gender~\cite{Skogli2013ADHD}, will be essential to strengthen the validity of findings and identify more robust factors underlying different adoption strategies.
Our results point to influences such as parents' existing management approaches and their level of overwhelm, but larger samples are needed to draw statistically sound conclusions.

Future work should also examine how \system{} can complement existing parent training programs and assess whether it could be integrated into such interventions.
In addition, incorporating children's perspectives—for example, through interviews as discussed in~\secref{sec:disc-children}—would provide a more holistic understanding of family dynamics. 

In addition, our current design intentionally limits parents' access to contextual information for privacy reasons.
However, we acknowledge that this decision may reduce parents' ability to fully understand the child's situation and could influence their trust in the system.
Future versions of \system{} could explore providing minimal, privacy-preserving context—such as simple summaries derived from steps, heart rate, or other aggregate signals—to help parents contextualize notifications while still maintaining strong protections for sensitive data.

Finally, technical improvements are necessary: (1) advancing from the current simple linear regression model (input: energy level, output: parental perception) to more sophisticated modeling approaches that may improve accuracy, and (2) enhancing networking capabilities, since current Apple Watch cellular models do not permit custom app installation, requiring reliance on participants' local WiFi connections.  

\section{Conclusion}

Families raising children with ADHD face daily challenges in sustaining constructive interactions, as children's hyperactive behaviors often disrupt routines and elevate parental stress.
This paper introduced \system{}, a wearable-sensing and notification probe designed to identify calm moments and prompt parents toward reflection and reinforcement in real time.
Through a four-week in-home study with 12 completed families, we compared notification strategies that varied in frequency and adaptivity, revealing how the timing and framing of prompts shaped parents' awareness, interpretation, and engagement.
While our findings demonstrated the feasibility of calm-moment detection and highlighted diverse interventions across families, they also underscored the importance of personalization and sensitivity to parents' existing stress and management strategies.
Together, these findings extend HCI work on sensing and adaptive notifications by highlighting both the technical feasibility and the lived complexities of integrating sensing-driven interventions into family routines in the context of hyperactivity.

\begin{acks}
This work was partially supported by NSF Award No. 2406099, NIMH Grant No. 1R44MH138145, and Carnegie Mellon University's Center for Machine Learning and Health. We thank the study participants and the anonymous reviewers for their valuable feedback.
\end{acks}

\bibliographystyle{ACM-Reference-Format}
\bibliography{paper}

\end{document}